\PassOptionsToPackage{pdfpagelabels=false}{hyperref}
\documentclass[fleqn,usenatbib]{mnras}

\usepackage{newtxtext,newtxmath}

\usepackage[T1]{fontenc}
\usepackage{ae,aecompl}

\usepackage{graphicx}	
\usepackage{amsmath}	
\usepackage{amssymb}	
\usepackage{bm}         
\usepackage{color}      
\usepackage{xspace}
\usepackage{epstopdf}

\newcommand{\dd}{{\rm d}}
\newcommand{\Fc}{{F_{\mathrm c}}}
\newcommand{\pt}{{p_{\mathrm t}}}
\newcommand{\pg}{{p_{\mathrm g}}}
\newcommand{\Kepler}{\textit{Kepler}\xspace}

\newcommand{\magenta}{}
\def\note #1]{{\bf\magenta #1]}}

\DeclareMathAlphabet{\mathcmr}{OT1}{cmr}{m}{it}
\newcommand{\cg}{\mathcmr{g}}

\usepackage{etoolbox}
\makeatletter
\patchcmd\@combinedblfloats{\box\@outputbox}{\unvbox\@outputbox}{}{\errmessage{\noexpand patch failed}}
\makeatother

\title[Damping rates and frequency corrections of \Kepler LEGACY stars]
      {Damping rates and frequency corrections of \Kepler LEGACY stars}

\author[G. Houdek et al.]{
G. Houdek,$^{1}$\thanks{E-mail: hg@phys.au.dk (GH)}
M. N. Lund,$^{1}$
R. Trampedach,$^{2,1}$
J. Christensen-Dalsgaard,$^{1}$
\newauthor\ R. Handberg$^{1}$
 and T. Appourchaux$^{3}$
\\
$^{1}$Stellar Astrophysics Centre, Department of Physics and Astronomy, Aarhus University, 8000 Aarhus C, DK\\
$^{2}$Space Science Institute, 4750 Walnut Street, Suite 205, Boulder, CO 80301, USA\\
$^{3}$Univ. Paris-Sud, Institut d'Astrophysique Spatiale, UMR 8617, CNRS, B\^{a}timent 121, 91405 Orsay Cedex, F
}

\date{Accepted XXX. Received YYY; in original form ZZZ}

\pubyear{2018}

\begin{document}
\label{firstpage}
\pagerange{\pageref{firstpage}--\pageref{lastpage}}
\maketitle

\begin{abstract}
Linear damping rates and modal frequency corrections of radial oscillation modes in
selected LEGACY main-sequence stars are
estimated by means of a nonadiabatic stability analysis. The selected stellar sample
covers stars observed by \Kepler with a large 
range of surface temperatures and surface gravities.
A nonlocal, time-dependent
convection model is perturbed to assess stability against pulsation modes. 
The mixing-length parameter is calibrated to the surface-convection-zone depth of 
a stellar model obtained
from fitting adiabatic frequencies to the LEGACY observations, and two of the 
nonlocal convection parameters are calibrated to the corresponding LEGACY
linewidth measurements. The remaining nonlocal convection parameters in the 
1D calculations are 
calibrated so as to reproduce profiles of turbulent pressure and 
of the anisotropy of the turbulent velocity field of corresponding 3D hydrodynamical
simulations. 
The atmospheric structure in the 1D stability analysis 
adopts a temperature-optical-depth relation derived from 3D hydrodynamical 
simulations.  
Despite the {small number of} parameters to adjust, 
we find good agreement with detailed shapes of both turbulent pressure
profiles and anisotropy profiles with depth, and with
damping rates as a function of frequency. 
Furthermore, we find the 
absolute modal frequency corrections, relative to a standard adiabatic 
pulsation calculation, 
to increase with surface temperature and surface gravity.
\end{abstract}

\begin{keywords}
Sun: oscillations -- convection -- hydrodynamics -- turbulence
\end{keywords}



\section{Introduction}

Global parameters of observed stellar targets are typically determined by fitting
stellar evolutionary models to spectroscopically determined surface values, such as
temperature, gravity, and chemical composition, together with minimizing the
differences between observed and adiabatically computed oscillation frequencies
by a nonlinear optimization approach 
\citep[e.g.][and references therein]{SilvaAguirreEtal17}.
However, in addition to the observed oscillation frequencies other, simultaneously
measured, pulsation properties,
such as linewidths and amplitudes of stochastically excited oscillations,
can be used 
to further constrain our stellar models.
The reliable determination of such additional mode parameters requires
high-quality seismic observations, such as 
from NASA's space mission \Kepler \citep{BoruckiEtal10, JCDEtal09b}. 
Moreover, with the recent success of three-dimensional (3D) hydrodynamical simulations of
outer stellar layers \citep[e.g.,][]{TrampedachEtal13, MagicEtal13,
LudwigSteffen16, KupkaEtal18},
it is now possible to use averaged simulation results for calibrating 
convection parameters of 1D stellar
models \citep[e.g.][]{LudwigEtal99, TrampedachEtal14b, MagicEtal15, Houdek17,
AarslevEtal18} or replacing the outer layers of 1D stellar models
by averaged 3D simulations for seismic analysis 
\citep[e.g.,][]{RosenthalEtal95, RosenthalEtal99, 
PiauEtal14, SonoiEtal15, SonoiEtal17, MagicWeiss16, BallEtal16, HoudekEtal17,
TrampedachEtal17, JoergensenEtal18, MosumgaardEtal18}.

First attempts to estimate mode damping of intrinsically stable, but stochastically
driven, pulsations in the Sun were reported by \citet{GoldreichKeeley77a}, who used 
a time-independent scalar turbulent viscosity to describe 
mode damping by turbulent convection. The first prediction of frequency-dependent solar
damping rates were reported by \citet{Gough80}, using his time-dependent,
local, convection formulation \citep{Gough77a}. \citet{JCDEtal89} compared Gough's
predictions with
linewidth measurements from the Big Bear Solar Observatory, and found qualitatively good
agreement. Quantitatively good agreement between solar linewidth measurements and 
damping-rate estimates was reported by \citet{Balmforth92a}, who used a nonlocal
generalization \citep{Gough77b} of Gough's local time-dependent convection formulation.
\citet{DupretEtal06a} also reported solar damping-rate calculations that were in good
agreement with solar linewidth measurements, using 
Grigahcene's\,et\,al.\,(\citeyear{GrigahceneEtal05}) generalization of 
Unno's\,(\citeyear{Unno67}) time-dependent convection model.

Predictions of frequency-dependent damping rates in more than 100 main-sequence 
stellar models were reported by \citet{HoudekEtal99}, who used
Gough's\,(\citeyear{Gough77b}, \citeyear{Gough77a}) nonlocal, time-dependent 
convection formulation,
as did  \citet{Balmforth92a} for the Sun,
in the nonadiabatic stability calculations. 
Frequency-dependent damping rates in red giants were addressed by, for example,
\citet{HoudekGough02}, \citet{DupretEtal09}, \citet{ GrosjeanEtal14} 
and \citet{AarslevEtal18}.
\citet{BelkacemEtal12} compared theoretical damping rates with
main-sequence and red-giant linewidth observations from CoRoT \citep{BaglinEtal09} 
and \Kepler at frequencies near the 
maximum oscillation power, $\nu_{\rm max}$.
Recent reviews of mode damping were published by
\citet{HoudekDupret15} and \citet[][see also \citealt{AarslevEtal18}]{SamadiEtal15}.

In this work we model frequency-dependent damping rates and modal frequency corrections 
for twelve selected LEGACY main-sequence stars that were observed by 
\Kepler \citep{Borucki16} and which
cover a substantial range of surface 
temperatures and surface gravities. The global 
stellar parameters and depths of surface convection zones are obtained from 
frequency-calibrated ASTEC evolutionary calculations \citep{SilvaAguirreEtal17}. 
{The ASTEC calculations adopted the classical mixing-length formulation of 
convection by \citet{BohmVitense58} with a fixed mixing-length parameter $\alpha=1.8$.
The mixing-length parameters in our model calculations are calibrated individually for 
each model to give the same convective envelope depths as the corresponding ASTEC models.}
Our adopted convection model by \citet{Gough77a} includes additional, nonlocal, 
parameters which need calibration. These additional convection parameters 
are calibrated against a grid of 3D hydrodynamical 
simulations by \citet{TrampedachEtal13} and LEGACY linewidth measurements 
by \citet{LundEtal17}.
The atmospheric 
structure in the 1D stability analysis 
adopts temperature-optical-depth relations 
derived from the 3D hydrodynamical 
simulations by \citet{TrampedachEtal14a}. 
Our theoretical, frequency-dependent damping rates, $\eta$, correspond to the 
half-width at half-maximum (HWHM) of the spectral peaks in the acoustic power 
spectrum, $P(\omega)$, where $\omega$ is the angular pulsation frequency.
The damping-rate calculations are similar to that by \citet{HoudekEtal17}, 
who applied it successfully to the solar case.

In addition to damping rates, our calculations also provide the modal frequency
corrections contributing to the 
so-called 'surface effects' \citep[e.g.][]{Brown84, Gough84,
Balmforth92b,RosenthalEtal99, Houdek10, GrigahceneEtal12}.  
The surface effects can be
divided into effects arising from the stellar mean structure, i.e. the structural effects,
as well as direct convective effects on pulsation modes, typically neglected
in standard adiabatic pulsation calculations, i.e. the modal effects 
\citep{RosenthalEtal95}. 
First attempts to assess the structural surface effects 
by means of so-called 'patched models', where the outer stellar layers of
a 1D model are replaced by an appropriately averaged 3D simulation, were reported by
\citet{RosenthalEtal95} and later by \citet{RosenthalEtal99}.
Structural effects were recently re-addressed by \citet{MagicWeiss16}, 
\citet{BallEtal16}, \citet{TrampedachEtal17}, and \citet{JoergensenEtal18},
and both structural and modal effects by
\citet{HoudekEtal17}, and \citet{SonoiEtal17}. 
Here we calculate the modal effects following the
procedure by \citet{HoudekEtal17}, including both nonadiabatic and convection-dynamical
effects associated with the pulsationally induced perturbations of the convective heat
(enthalpy) flux and momentum flux (turbulent pressure).
In Section~\ref{s:legacy-sample} we summarize the LEGACY pipeline and the
stellar evolutionary computations of the frequency-calibrated LEGACY models.
Section~\ref{s:3D-simulations} summarizes the details of our adopted grid of
3D convection simulations, and Section~\ref{s:stability} addresses the 1D 
stability calculations.
Results are presented in Section~\ref{s:results} and followed by a discussion in 
our final Section~\ref{s:discussion}.

\section{LEGACY stellar sample}
\label{s:legacy-sample}
In the LEGACY analysis \citep{LundEtal17} parameters of the solar-like oscillation modes
were extracted by fitting a global model to the power-density spectrum -- this model
includes both the oscillation modes and the signature from the stellar granulation
background. Each oscillation mode is modelled by a Lorentzian function
\begin{equation}
L_{nlm}(\nu) = \frac{E_{lm}(i_{\star})\, \tilde{V}^2_l\, S_{n0}}{1 + 4\Gamma_{nl}^{-2}(\nu - \nu_{nl} + m\nu_s)^2}\, .
\end{equation}

The mode with radial order $n$ and spherical degree $l$ is characterized 
by a central frequency $\nu_{nl}$ of the zonal component $m=0$
(azimuthal order), the height $S_{n0}$, which is modulated by a mode visibility $\tilde{V}^2_l$
from spatial filtering, the geometrical relative visibility $E_{lm}(i_{\star})$
between azimuthal components from the stellar inclination $i_{\star}$, and by 
the linewidth $\Gamma_{nl}$, which is the full width at half maximum (FWHM) of the fitted
Lorentzian to the spectral peaks in the observed oscillation power spectrum. In the
presence of rotation the mode will further be split into a multiplet from the rotational
advection, $\nu_s$.

In the fitting, which is performed in a Bayesian framework using a Markov chain Monte Carlo
sampler \citep{emcee}, the linewidths and amplitudes of the radial modes are optimized. 
Corresponding values for nonradial modes are obtained from a linear interpolation to the 
radial modes. In the fitting 
it is furthermore the mode amplitude rather than the height that is optimized
\citep{LundEtal17}, and from 
this and the linewidth the height that goes into the global model is computed.

In the analysis of the linewidths reference stellar evolution models were
obtained from the LEGACY fits to the observed frequencies, combined with `classical' 
spectroscopic observations, carried out by \citet{SilvaAguirreEtal17}.
Specifically, except in one case we used the results of the ASTFIT procedure.
The models were computed with the ASTEC evolution code
\citep{JCD08}, using OPAL opacity tables \citep{IglesiasRogers96}
with the \citet{grevesse_noels93} composition, the OPAL equation of state
\citep{RogersNayfonov02} and the \citet{Anguloetal99} nuclear reaction rates,
with the \citet{Imbrianietal05} update to the ${}^{14}{\rm N}$ rate.
Convection was treated with the \citet{BohmVitense58} mixing-length
formulation, in all cases using a mixing-length parameter
$\alpha_{\rm ML} = 1.8$, roughly corresponding to solar calibration.
For models with mass $M \le 1.16 \,{\rm M}_\odot$,
${\rm M}_\odot$ being the solar mass, diffusion and settling 
were included using the \citet{Michaud_Proffitt_93} approximation.
Adiabatic oscillation frequencies were computed with the ADIPLS \citep{JCD08adipls} code;
the frequencies were corrected for surface effects using a scaled solar
surface correction \citep{JCD12}.
The {ASTFIT analysis} used a grid with a step of $0.01\, {\rm M}_\odot$.
The fitting procedure was described in some detail by \citet{Gillilandetal13}
and \citet{SilvaAguirreEtal15}.
Briefly, the best-fitting model was identified along each of a suitable
subset of the evolution tracks, and the model parameters were obtained
by computing a likelihood-weighted average and standard deviation
based on these best-fitting models.

The so obtained model properties of our twelve selected LEGACY stars are listed
in Table~\ref{t:legacy-sample}.

\section{The 3D hydrodynamical simulation grid}
\label{s:3D-simulations}

We employ the grid of 3D hydrodynamical simulations of 37 deep convective
stellar atmospheres (including a solar one) by \citet{TrampedachEtal13}.
The simulations evolve the conservation equations for mass, momentum and
energy on a regular grid, optimized in the vertical direction to capture
the photospheric transition.
Radiative transfer is solved explicitly with the hydrodynamics, and
line-blanketing (non-greyness) is accounted for by a binning of the
monochromatic opacities, as developed by \citet{Nordlund82}.
The monochromatic opacities are detailed in \citet{TrampedachEtal14a}
and the thermodynamics is supplied by the \citet{MihalasEtal88}
equation of state, custom computed for the exact same 15-element
mixture as the opacities. The grid is computed for a solar composition
with heavy-element abundance by mass of $Z=0.018055$, and helium abundance
by mass of $Y=0.24500$.
Horizontal boundaries are cyclic and the top and bottom boundaries are
open and transmitting, minimizing their effect on the interior of the
simulation. The constant entropy assigned to the inflows at the bottom is
adjusted to obtain the desired effective temperatures, $T_{\rm eff}$.

\section{Stability computations}
\label{s:stability}

The coupling between convection and oscillation, in particular the pulsationally induced
perturbations to both the convective heat (enthalpy) 
flux and momentum flux (turbulent pressure),
requires a time-dependent formulation of convection. 
A meaningful assessment of the physics implemented in such a convection formulation 
requires a self-consistent implementation in both the equilibrium stellar 
structure and stability analysis within intended limits, such as the assumption 
of a Boussinesq fluid. 
Furthermore, a consistent
inclusion of turbulent pressure in a 1D equilibrium model is only possible
within the framework of a nonlocal formulation of convection \citep{Gough77a}, which
avoids the otherwise present singularities at the convective boundaries in the stellar
structure equations.
Here we adopt the convection model by \citet{Gough77b, Gough77a}, which was recently
reviewed in detail by \citet[][see also \citealt{AarslevEtal18}]{HoudekDupret15}.
We therefore introduce here only the main properties of Gough's nonlocal convection
formulation.

In addition to avoiding singularities in the stellar structure equations when
turbulent pressure is included in the computations, a nonlocal convection formulation 
also helps to suppress the unphysically rapid spatial oscillations of the pulsation
eigenfunctions in the deep, nearly adiabatically stratified, stellar layers, where 
resulting cancellation effects can lead to erroneous work integrals and {resulting} 
damping rates \citep[e.g.,][]{BakerGough79, HoudekDupret15}.
Additional nonlocal properties are that (1) convective eddies sample the superadiabatic
temperature gradient
$\beta:=-\dd T/\dd r\,+\,(\dd T/\dd r)_s$, 
where $T$ is temperature, $r$ is radius and $s$ is specific entropy,
over a vertical layer with an extent $\ell$ that is typically 
of the order of a pressure scale height, and (2) the turbulent fluxes of 
heat and momentum are determined not only by the local conditions at a point $r_0$ in the star, 
but from an (weighted) average of all convective eddies that are centred within this
vertical stellar layer of extent $\ell$.

\begin{table*}
	\centering
	\caption{LEGACY model parameters. The first eight columns show the  
	 Kepler Input Catalog (KIC) number and global model parameters as 
	 determined from fitting observed to adiabatic model frequencies
	 \citep{SilvaAguirreEtal17, LundEtal17}, 
	 i.e. surface temperature, $T_{\rm eff}$, 
	 and logarithmic gravity, $\log\cg$, the surface abundances by mass of 
	 heavy elements, $Z$, and helium, $Y$,
	 stellar mass, $M$, and surface luminosity, $L$, in solar units, and the depth 
	 of the surface	convection zone, $d_{\rm c}$, in units of stellar radius $R_\star$. 
	 The remaining six columns are the calibrated convection parameters
	 adopted in our stability calculations, i.e. the nonlocal parameters, $a$, $b$ 
	 and $c$, the mixing-length	parameter, $\alpha$, and the anisotropy 
	 parameters, $\varPhi_{\rm s}$ and $\varPhi_{\rm c}$, at the stellar
	 surface and in the deep interior, respectively.
     }
	\label{t:legacy-sample}
	\begin{tabular}{rcccccccccrcrc} 
		\hline
		KIC & {$T_{\rm eff}\,$(K)} & $\log \cg$ (cgs) & $Z$ & $Y$ & $M/{\rm M}_\odot$ & $L/{\rm L}_\odot$ & $d_{\rm c}/R_\star$ & $a$ & $b$ & $c$ & $\alpha $ & $ \varPhi_{\rm s} $ & $ \varPhi_{\rm c}$\\
		\hline
		6933899 & 5888 & 4.087 & 0.0249 & 0.2830 & 1.13 & 2.731 & 0.287 & 17.89& 44.72&13.42& 2.12 & 10.05 & 2.05 \\
		10079226& 5948 & 4.365 & 0.0266 & 0.2853 & 1.11 & 1.474 & 0.262 & 12.25& 44.72& 6.86& 2.14 & 10.50 & 1.50 \\
		10516096& 5983 & 4.177 & 0.0193 & 0.2674 & 1.11 & 2.325 & 0.272 & 15.81& 89.44&10.95& 2.16 &  9.90 & 1.90 \\
		6116048 & 6031 & 4.275 & 0.0156 & 0.2635 & 1.06 & 1.832 & 0.263 & 12.04& 89.44& 7.87& 2.19 & 13.67 & 1.67 \\
		8379927 & 6112 & 4.391 & 0.0215 & 0.2696 & 1.16 & 1.617 & 0.224 & 14.14& 89.44& 7.07& 2.13 & 10.00 & 1.50 \\
		12009504& 6217 & 4.214 & 0.0174 & 0.2653 & 1.19 & 2.670 & 0.202 & 17.32& 89.44&14.14& 2.15 &  9.83 & 1.83 \\
		6225718 & 6287 & 4.318 & 0.0174 & 0.2653 & 1.17 & 2.159 & 0.187 & 17.32& 89.44&11.62& 2.14 &  9.67 & 1.67 \\
		1435467 & 6334 & 4.106 & 0.0215 & 0.2696 & 1.38 & 4.273 & 0.157 & 89.44& 89.44&22.36& 2.11 & 10.17 & 1.67 \\
		9139163 & 6397 & 4.190 & 0.0295 & 0.2776 & 1.40 & 3.717 & 0.143 & 89.44& 89.44&29.33& 2.12 &  8.97 & 1.67 \\
		9206432 & 6582 & 4.224 & 0.0266 & 0.2746 & 1.42 & 3.910 & 0.108 & 89.44& 89.44&15.49& 2.19 &  7.20 & 2.00 \\
		2837475 & 6681 & 4.162 & 0.0193 & 0.2674 & 1.42 & 4.786 & 0.085 & 89.44& 89.44&15.81& 2.20 &  7.00 & 2.00 \\
		11253226& 6715 & 4.171 & 0.0174 & 0.2653 & 1.40 & 4.723 & 0.080 & 89.44& 89.44&20.00& 2.23 &  7.40 & 2.20 \\
		\hline
	\end{tabular}
\end{table*}

Gough's\,(\citeyear{Gough77b}) concept of averaging over convective eddies is based on 
Spiegel's\,(\citeyear{Spiegel63}) finding that the linearized equations of motions for
determining the convective growth rate $\sigma$ of a convective eddy, i.e. the rate with
which the convective fluctuations grow with time, satisfy a variational equation for 
$\sigma$, if the locally 
defined superadiabatic temperature gradient $\beta(r_0)$ is replaced by the 
averaged, i.e. nonlocal, value
\citep{Gough77b, Gough77a}
\begin{equation}
{\cal B}(r)=\frac{2}{\ell}\int\beta(r_0)\cos^2\left[\pi(r_0-r)/\ell\right]\,\dd r_0\,.
\label{e:averaged-beta}
\end{equation}
In expression~(\ref{e:averaged-beta}) the averaging is obtained by taking account of
contributions from convective eddies centred at $r_0$ and the range of integration is 
the vertical extent $\ell$, which scales with the local pressure scale height.
Based on Spiegel's finding on the convective {growth} rate, \citet{Gough77b} 
suggested similar expressions for the
averaged, nonlocal, convective heat, ${\cal\Fc}$, and momentum, $\mathcal{P}_{\rm t}$,
fluxes, with $\beta(r_0)$ in Eq.~(\ref{e:averaged-beta}) being replaced by the locally
computed convective fluxes $\Fc(r_0)$ and $\pt(r_0)$, respectively.
This would, however, lead to a system of integro-differential equations for the stellar
structure, which would be difficult to solve numerically.
\citet{Gough77b} suggested therefore to approximate the kernel 
${\cal K}:=2\cos^2[\pi(r_0-r)/\ell]$ in Eq.~(\ref{e:averaged-beta}) by the expression
${\cal K}\simeq a\exp(-a|r_0-r|/\ell)/2$, and setting the integration limits formally
to $\pm\infty$, which leads to an integral expression that represents the solution 
to the $2^{\rm nd}$-order
differential equation 
\begin{equation}
\frac{\dd^2{\cal\Fc}}{\dd\ln p^2}=\frac{a^2}{\alpha^2}\left({\cal\Fc}-\Fc\right)\,
\end{equation}
\citep{Gough77b, Balmforth92a, HoudekDupret15}
for the nonlocal convective heat flux, ${\cal\Fc}$, for example,
where $\ln p$ ($p$ is total pressure) is now the new independent depth variable
(as implemented in the calculations), 
$\alpha:=-\ell{\rm d}\ln p/{\rm d} r$ is the mixing-length parameter, and $a$
is another dimensionless, nonlocal, parameter. Similar expressions are obtained for the
averaged, nonlocal, superadiabatic temperature gradient, $\cal B$, and turbulent pressure, 
$\cal P_{\mathrm t}$, introducing the remaining nonlocal (dimensionless) parameters $b$ and
$c$, respectively.
These nonlocal parameters control the spatial coherence of the ensemble of eddies 
contributing to the total heat ($a$) and momentum ($c$) fluxes, and the degree to which
the turbulent fluxes are coupled to the local stratification ($b$).
Roughly speaking, the parameters $a$, $b$ and $c$ control the
degree of `non-locality' of convection; low values imply highly nonlocal
solutions, and in the limit $a, b, c\rightarrow\infty$ the system of equations
formally reduces to the local formulation ({except near the boundaries
of convection zones}, where the local equations are singular).
Theoretical values for the dimensionless, nonlocal, parameters $a$, $b$ and $c$ can 
be obtained by demanding that the terms in a Taylor expansion about $r$ of the 
exact, $2\cos^2[\pi(r_0-r)/\ell]$, and approximate, $a\exp(-a|r_0-r|/\ell)/2$,  kernels
differ only at fourth order, {resulting in a theoretical estimate for} 
$a\simeq 7.8$ \citep{Gough77b}.
{The standard mixing-length assumption of assigning
a unique scale to turbulent eddies at any given location can, however,  
cause too much smoothing, leading to larger values for $a, b$ or $c$
\citep{HoudekGough02}.}
Therefore, values for $a, b$ and $c$, which typically differ, need
to be determined from calibration in a similar way as it is common for the mixing-length 
parameter, $\alpha$, in stellar evolutionary theory\footnote{We should remain aware that 
calibrated values for the mixing-length parameter, $\alpha$, in stellar evolutionary
calculations can be as large as $\alpha\gtrsim 2$ (see also Table\,\ref{t:legacy-sample}),
which is almost one magnitude larger than what the underlying assumptions, e.g., the
Boussinesq approximation \citep{SpiegelVeronis60}, for a mixing-length model would allow
\citep[e.g.][and references therein]{GoughWeiss76}.}.
It should be mentioned that our adopted nonlocal convection formulation does not treat 
the overshoot regions correctly, where
the nonlocal enthalpy flux, $\cal\Fc$, remains positive although it should be negative.
The overshoot regions may therefore not be suitable 
for calibrating the nonlocal parameters to 3D simulation results
(see also discussion in Section~\ref{ss:varpar}).
The effect of such positive $\cal\Fc$ values in the 
overshoot regions on the damping rates and 
eigenfrequencies is, however, negligible.

Another parameter in Gough's\,(\citeyear{Gough77b}) time-dependent convection model 
controls the anisotropy, $\varPhi:=\overline{\bm{u}\cdot\bm{u}}/\overline{w^2}$
(overbars indicate horizontal averages), of the convective velocity field $\bm{u}=(u,v,w)$.
This parameter enters as a multiplicative factor in the inertia term of the 
linearized momentum equation for the convective fluctuations, thereby effectively
increasing the inertia of the vertically moving convective eddies as a result of the
coupling between vertical 
and horizontal motion. For a solenoidal convective velocity field, it can be related 
to the shape of the convective eddies in the sense that $\varPhi\rightarrow 1$ represents 
thin, needle-like eddies. Larger $\varPhi$ values increase the eddie's inertia, thereby 
describing the diversion of the vertical motion into horizontal flows as a reduction 
of the convective efficacy. 
We describe the depth dependence of the anisotropy parameter $\varPhi$
with an analytical function guided by the 3D solution and by adopting approximately the 
maximum 3D value, $\varPhi_{\rm s}$, in the, convectively inefficient, surface layers 
and the minimum 3D value, $\varPhi_{\rm c}$, in the deep, convectively efficient, layers
in our 1D convection model.

The stability computations are carried out as in \citet{HoudekEtal17}, 
but with the global stellar parameters adopted from \citet{SilvaAguirreEtal17}%
\footnote{The parameters for KIC 6933899 were obtained from the BASTA
fit, whereas in the remaining cases ASTFIT results
(see Section~\ref{s:legacy-sample}) were used.}.
These parameters are listed 
in Table~\ref{t:legacy-sample}, together with the abundances by
mass of helium, $Y$, and heavy elements, $Z$, as obtained from minimizing the
differences between adiabatically computed
oscillation frequencies and {\it Kepler} observations by least squares 
(see Section~\ref{s:legacy-sample}).
The mixing length was calibrated to within 1.5\% of the convection-zone 
depths, $d_{\rm c}/{R_\star}$ (also listed in Table~\ref{t:legacy-sample}), 
of the calibrated ASTEC evolutionary models.
Both the envelope and pulsation calculations assume the generalized Eddington
approximation to radiative transfer \citep{UnnoSpiegel66}.
The opacities are obtained from the OPAL tables
\citep{IglesiasRogers96}, supplemented at low temperature by tables 
from \citet{Kurucz91}.
The EOS includes a detailed treatment of the ionization
of H, He, C, N, and O, and a treatment of the first ionization of the next
seven most abundant elements \citep{JCD82}.
The integration of stellar-structure equations starts 
at an optical depth of $\tau=10^{-4}$ and ends at a radius fraction $r/R_\star=0.2$.
The temperature gradient in the plane-parallel atmosphere is corrected by 
using a radially varying Eddington factor obtained from interpolating 
in the {$T(\tau)$-relations} 
($\tau$ is optical depth) of
Trampedach's\,et\,al.\,(\citeyear{TrampedachEtal14a}) 
{grid of 3D hydrodynamical simulations}.

\begin{figure*}
	\includegraphics[width=0.92\textwidth]{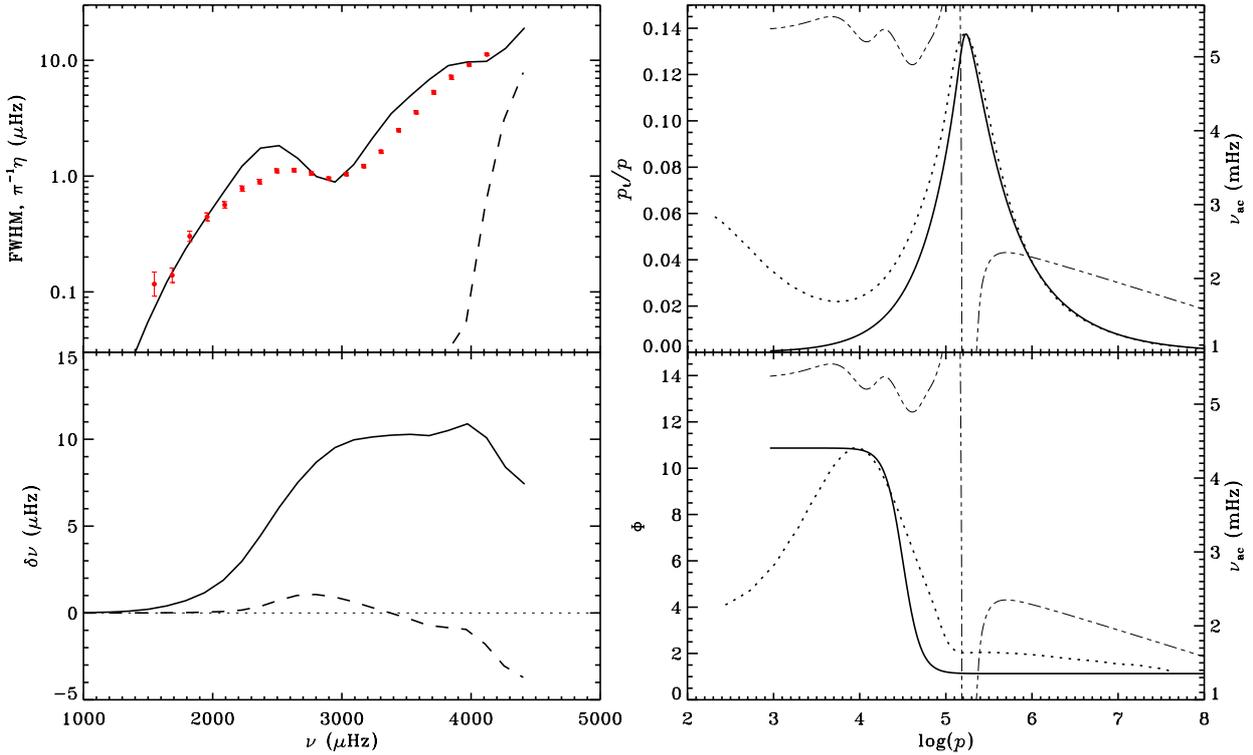}
    \caption{Solar model properties calculated by \citet{HoudekEtal17}, who adopted 
    the nonlocal parameters $a=8.37$, $b=44.72$, $c=4.43$, $\varPhi_{\rm c}=1.13$ for
    the anisotropy parameter in the deep convective interior, and $Y=0.245$ and
     $Z=0.018055$ for the helium and heavy-element abundances by mass.. 
    Top left: comparison of BiSON linewidths   
    \citep[symbols with error bars from][]{ChaplinEtal05} 
     {and $\pi^{-1} \eta$,
     where $\eta$ is the theoretical damping rate for radial modes;
     note that $\pi^{-1} \eta$ corresponds to the full width at half maximum,
     in cyclic frequency, of the peaks in the observed power spectrum.}
    The solid curve is the result of the full
     calculation in which 
    both the perturbations to the convective heat and momentum fluxes were included. 
    The dashed curve displays the 
    damping rates for a stability calculation in which the perturbation to the momentum
     flux, $\updelta p_{\rm t}$, was
    suppressed. 
    Bottom left: modal frequency corrections, $\updelta\nu$, relative to a standard
     adiabatic frequency calculation
    {in which the Lagrangian perturbation to turbulent pressure vanishes.}
    The solid curve includes both the perturbations to the
     convective heat and momentum fluxes
    and the dashed curve is the modal frequency correction for a calculation 
    in which the perturbation to the momentum flux, $\updelta p_{\rm t}$, was suppressed. 
    Top right: comparison of the turbulent pressure profile, $p_{\rm t}/p$, between 
    the 3D simulation (dotted curve; \citealt{TrampedachEtal14b})
    and the 1D nonlocal solar envelope model (solid curve). 
    Bottom right: comparison of 
    the turbulent velocity anisotropy, $\varPhi$,
    between the 3D simulation (dotted curve) and the 1D model (solid curve). 
     {In the right-hand panels the {double}-dot-dashed curve 
     is the acoustic cutoff frequency $\nu_{\rm ac}$ 
     \citep{DeubnerGough84} of the 1D, nonlocal, reference model}.} 
    \label{f:sun}
\end{figure*}

The linear nonadiabatic pulsation calculations are carried out using
the same nonlocal convection formulation with the assumption that all eddies in the cascade
respond to the pulsation in phase with the dominant large eddies.
A simple thermal outer boundary condition is adopted at the temperature minimum 
where for the mechanical boundary condition the solutions are 
matched smoothly onto those of a plane-parallel isothermal atmosphere for frequencies 
below the acoustic cutoff frequency {$\nu_{\rm ac}$}, i.e. the solutions are matched
onto an exponential solution with the energy density decreasing with height in the
atmospheres. For modes with frequencies larger than {$\nu_{\rm ac}$} the eigenfunctions 
are matched to an outwardly running wave, allowing for transmission of energy into the
atmosphere \citep[e.g.,][]{BalmforthGough90, BalmforthEtal01}.
At the base of the model envelope ($r/R_\star=0.2$) the 
conditions of adiabaticity and vanishing of the displacement eigenfunction 
are imposed. Only radial p modes are considered.

In addition to the fully nonadiabatic calculations, which include the modal
perturbations to the convective heat and momentum fluxes, we
also computed frequencies in the adiabatic approximation for the same equilibrium models.
In our adiabatic calculation we assume that the modal Lagrangian perturbation to the
turbulent pressure, $\updelta\pt$, is in quadrature with the modal density
perturbation, $\updelta\rho$. This leads to a modification of the first adiabatic 
exponent, $\gamma_1:=(\partial\ln\pg/\partial\ln\rho)_s$ 
($\pg$ is gas pressure), in the linearized, adiabatic
pulsation equations, i.e. the Lagrangian (total) pressure perturbation, 
$\updelta\ln p=(\pg/p)\gamma_1\updelta\ln\rho$ \citep{HoudekEtal17}, also referred 
to as the {\it reduced} $\gamma_1$ \citep{RosenthalEtal95}.
From taking the differences between the fully nonadiabatic, $\nu_{\rm na}$, and 
adiabatic, $\nu_{\rm a}$, frequencies, we obtain estimates for the 
modal frequency corrections, $\delta\nu:=\nu_{\rm na}-\nu_{\rm a}$, i.e. the modal
contribution to the so-called ``surface effects'' 
\citep[e.g.,][]{RosenthalEtal99}.

\section{Results}
\label{s:results}
\subsection{The solar benchmark case}
We start by reviewing Houdek's\,et\,al.\,(\citeyear{HoudekEtal17}) solar results
depicted in Fig.\,\ref{f:sun}. 
The adopted 1D profile for the velocity-anisotropy parameter $\varPhi$ 
(solid curve in the lower right panel) is in good
agreement with the 3D profile (dotted curve), particularly in the mode-propagating 
layers ($\log p\geq 5.2$).
The regions
where the modes are either propagating ($\log p\geq 5.2$) or evanescent ($\log p<5.2$)
are indicated by the acoustic cutoff frequency, $\nu_{\rm ac}$ 
(double-dot-dashed curve)\footnote{The acoustic cutoff frequency, $\nu_{\rm ac}$, 
in figure~4 of \citet{HoudekEtal17} is erroneously shifted in depth.}. 
The mode frequencies are predominantly determined by the mode-propagating regions which
therefore also affect the modal frequency corrections discussed later in
Section~\ref{s:discussion}.
The remaining differences in $\varPhi$ with respect to the 3D solution
in the mode-propagating layers do, however, 
affect the solutions of our damping-rate calculations. We note that the effect 
of $\varPhi$ on mode damping is strongly coupled to the filling factor 
(fractional area of, e.g., upflows), 
which is always $1/2$
in any 1D convection model that adopts 
the Boussinesq approximation \citep{SpiegelVeronis60} and assumes
symmetry between up- and downflows. 
Therefore, it is not expected that our 1D convection 
treatment yields the observed damping rates, even if the complete 3D $\varPhi$-profile 
is adopted in our stability analyses. Only a full 3D hydrodynamical simulation 
of the full oscillation properties would adequately describe the problem. 
This is still out of reach as of today. \citet{HoudekEtal17} therefore found it 
more useful and 
relevant to reproduce the observed damping rates by adopting the 1D
$\varPhi$-profile as indicated by the solid curve in the lower right panel of 
Fig.~\ref{f:sun} for the Sun.
We do adopt the same procedure here for our LEGACY models.
Moreover, the inclusion of a depth-dependent 
$\varPhi$-profile in our 1D pulsation analyses, with a shape and extrema adopted 
from 3D simulations, and its consistent pulsational perturbation, $\updelta\varPhi$ 
(the Lagrangian operator $\updelta$ indicates a perturbation following the motion), 
in the stability analyses, is a novelty in this work.
The expression for $\updelta\varPhi$ is given in 
e.g., \citet[][equation 74 and Appendix B]{HoudekDupret15}.
The turbulent pressure, $p_{\rm t}$, on the other hand, is robust to changes in filling 
factor, since it depends on the unsigned vertical convective velocity, and  
therefore reproduces the 3D solution very well in the mode-propagating 
layers.
In the outer stellar layers, where the differences in the $p_{\rm t}$-profiles 
between the 1D and 3D solutions are largest, the 
contribution of the perturbed turbulent pressure, $\updelta\pt$, to the damping 
rates, $\eta$,
is negligible as suggested by an analysis of the associated work integral $W_{\rm t}$
\citep[see, e.g.,][]{Balmforth92a, Houdek96}.
Only at deeper stellar layers with $\log p\gtrsim 5.0$ does the work of $\updelta\pt$ start
to become relevant with a relative contribution of about 8\% to $W_{\rm t}$ for a mode with
a frequency close to $\nu_{\rm max}$.
At these deeper stellar layers, however, the differences in the $p_{\rm t}$-profiles 
between the 1D and 3D solutions are already small.
Similar results are also observed for our LEGACY models.

The dashed curves in the left panels of Fig.\,\ref{f:sun} show the
(positive) damping rates (top panel) and modal frequency corrections (bottom panel)
for a computation in which the pulsationally perturbed turbulent 
pressure, $\updelta\pt$, is artificially suppressed. As expected, only the 
highest {frequency} modes are found to be stable, mainly
due to the radiative damping of these high-frequency modes, whereas for
frequencies less than about 3.8\,mHz the pulsation modes are predicted to be unstable 
in this case \citep[see also e.g.][]{HoudekEtal99}.

Modal frequency corrections for a computation in which
$\updelta\pt$ is artificially suppressed (dashed curve in bottom left panel)
indicate that nonadiabatic effects from the pulsationally perturbed 
radiative and convective heat fluxes are rather minute, i.e. most of the modal 
frequency corrections as indicated by the solid curve in the bottom left panel is due 
to the effects of convection dynamics associated with $\updelta\pt$.
We should, however, remain aware that due to the nonlinear nature of the
convection-pulsation interaction, the individual contributions to the modal surface effects
cannot simply be explained by linear superposition, 
e.g., by simply taking the differences
between the solid and dashed curves in the left panels of Fig.~\ref{f:sun}.  

\subsection{Effects of varying model parameters}
\label{ss:varpar}

\begin{figure*}
	\includegraphics[width=0.92\textwidth]{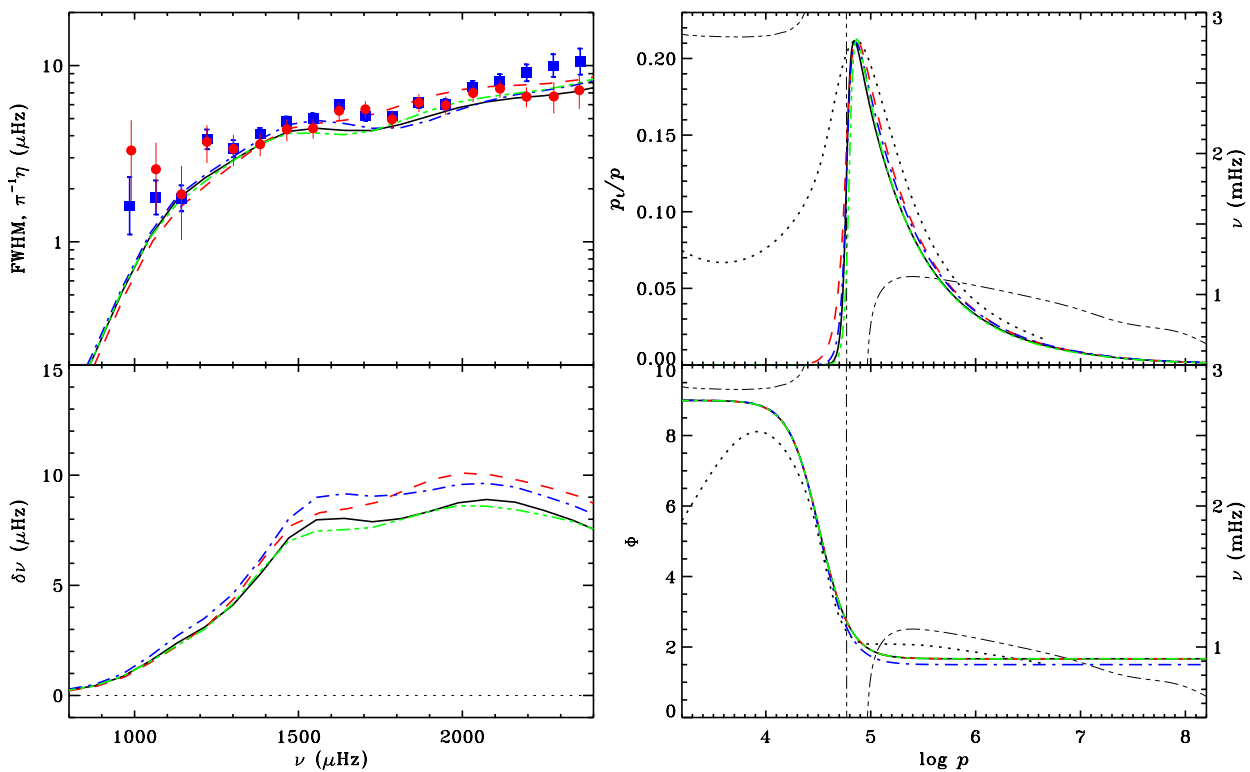}
    \caption{Model properties of KIC\;9139163. Results are shown for four different models
     with only one model parameter being different in each model calculation: 
     the solid, black, curves are the results for 
     the model parameters listed in Table~\ref{t:legacy-sample}, i.e. the 
     reference model; the dashed, red, curves are the
     results for a model computation in which the depth of the surface convection zone 
     is increased by 5\%, {relatively to the ASTEC model}; 
     the dot-dashed, blue, curves assume, in the deep 
     convection layers, the smaller velocity-anisotropy parameter $\varPhi_{\rm c}=3/2$, 
     relatively to the reference value $\varPhi_{\rm c}=5/3$; the
     {double}-dot-dashed, green, curves adopt the heavy-element abundance by mass of the 
     3D grid, $Z=0.018055$, and helium abundance by mass, $Y=0.245$.  
    Top left: comparison of observed linewidths (red solid circles with error bars 
    are the LEGACY data from
     \citealt{LundEtal17}, blue filled squares with error bars are from
     \citealt{AppourchauxEtal14, AppourchauxEtalCorr16})   
     and $\pi^{-1} \eta$ (curves).
    Bottom left: radial modal frequency
    corrections $\updelta\nu$ relative to a standard adiabatic pulsation calculation.
    Top right: comparison of the turbulent pressure profile 
     $p_{\rm t}/p$ between 3D simulation (dotted curve; \citealt{TrampedachEtal14a})
     and the four 1D nonlocal stellar envelope models (remaining curves). 
    Bottom right: comparison of the turbulent velocity anisotropy $\varPhi$
     between 3D simulation (dotted curve) and 1D models (remaining curves). 
     The acouctic cutoff frequency, $\nu_{\rm ac}$, of the reference model is 
     indicated by the {double}-dot-dashed, black, curves in the right panels.      
     See also caption to Fig.~\ref{f:sun}.
     }
    \label{f:punto}
\end{figure*}

The effect of varying the nonlocal parameters $a$, $b$ and $c$ on the estimated damping
rates were presented in detail
by \citet{Balmforth92a} for the solar case and recently by \citet{AarslevEtal18} for
red-giant stars. We therefore only summarize their effects here for the LEGACY stars, but
discuss additionally the effects of varying the depth of the surface convection 
zone, $d_{\rm c}$, velocity anisotropy, $\varPhi_{\rm c}$, and heavy-element abundance 
$Z$, on our results for the LEGACY star KIC\,9139163, depicted in Fig.~\ref{f:punto}.

The best-constrained nonlocal convection parameter is $c$, which controls the degree of 
``nonlocality'' of the turbulent pressure, $\pt$, i.e. the spatial 
coherence of the ensemble of eddies contributing to the turbulent momentum flux.
We calibrate $c$ by making the maximum value of $\pt$ in the superadiabatic boundary
layer of the 1D solution of our reference model (solid curve in the upper right 
panel of Fig.~\ref{f:punto}) agree with the value obtained from the 3D simulation result 
(dotted curve in the upper right panel). The
stellar depth at which these maxima occur agree well between the 1D and 3D
solutions without any further calibration, provided the heavy-element abundance is 
similar between the 1D and 3D models (see {double}-dot-dashed, green, curve in upper
right panel, and discussion below).
Similarly to $c$, the role of the nonlocal parameter $a$ is to control the
spatial coherence of the ensemble of eddies contributing to the convective
heat (enthalpy) flux. This parameter cannot be constrained in a straightforward manner
by the 3D simulation, because the adopted nonlocal convection model does not provide
the correct enthalpy flux in the overshoot regions. Instead we calibrate the
nonlocal parameter $a$ so as to obtain a good agreement between observed linewidths
and computed damping rates (upper left panel of Fig.~\ref{f:punto}).  
Larger values of $a$ result in a less pronounced depression in the damping rates, $\eta$,
near $\nu_{\rm max}\simeq1730\,\mu$Hz (see also \citealt{AarslevEtal18}).
The third nonlocal convection parameter $b$ controls the
degree to which the turbulent fluxes are coupled to the local stratification.
The larger the value of $b$
the stronger the turbulent fluxes are coupled to the local stratification,
resulting in less drastic changes in the stratification such as the superadiabatic 
temperature gradient (see also \citealt{Balmforth92a} and \citealt{AarslevEtal18}).
Similarly to $a$, changes to the value of $b$ affects predominantly the amount of
the depression in the damping rate, $\eta$, near $\nu_{\rm max}$, leading to some
degeneracy between $a$ and $b$, and consequently to a correlation between the results of
calibrating $a$ and $b$ to the observed linewidths. A possible remedy could include a more
extensive use of 3D simulations by calibrating $b$ to the super\-adiabatic 
temperature gradient
of 3D simulation results thereby breaking the degeneracy between $a$ and $b$, which we
plan in a future effort to improve the convection model.

In Fig.~\ref{f:punto} we show for KIC\,9139163 the effect of reducing the 
velocity-anisotropy 
parameter, $\varPhi_{\rm c}$ in the deep convection zone (lower right panel) from
$\varPhi_{\rm c}=5/3$ (solid, black, curve; 
overplotted by identical dashed, red, and {double}-dot-dashed, green, curves) 
to $\varPhi_{\rm c}=3/2$ 
(dot-dashed, blue, curve) keeping, however, the other model properties constant, including
the depth of the convection zone, $d_{\rm c}$, $\max(\pt)$ and $\varPhi_{\rm s}$. 
The corresponding results in the other panels are also depicted by dot-dashed, blue, 
curves and should be compared with the reference solutions illustrated by the 
solid, black, curves.
There is rather little effect on the turbulent pressure profile (upper right panel) but
noticeable effect on the damping rates (upper left panel) and modal surface correction
(lower left panel).
Reducing $\varPhi_{\rm c}$ decreases the characteristic 
timescale of the convection and consequently increases the frequency at which energy is 
exchanged most effectively between convection and pulsation. This is demonstrated 
in the upper left panel of Fig.~\ref{f:punto} by the increased frequency 
at which the minimum in the depression of the damping rates is observed 
(dot-dashed, blue, curve). 
Varying the value of $\varPhi_{\rm s}$ has
{comparatively} little effect on $\eta$ and 
the surface corrections (not shown here). Also, different
3D simulation codes predict different values for $\varPhi_{\rm s}$ 
(F. Kupka, personal communication, see also \citealt{KupkaMuthsam17}), possibly 
because of different surface boundary conditions in the 
various 3D simulation codes.

The effect of increasing the depth of the convection zone, $d_{\rm c}$, by 5\%, relatively
to the ASTEC evolutionary model, is illustrated by the dashed, red, curves in 
Fig.~\ref{f:punto}. The results
should be compared with the reference model (solid, black, curves). There is a 
slightly better agreement of the $\pt$-profile with the 3D solution (dotted, black, curve),
depicted in the upper right panel. The damping-rate increase with $d_{\rm c}$ 
(upper left panel), however, is substantial, as is the increase of the modal frequency
correction (lower left panel). A deeper surface convection zone requires a smaller
value for the calibrated nonlocal convection parameter $c$, explaining in part the
better agreement of the $\pt$-profile with the 3D simulation results, as a result of a 
slower exponential decay of $\pt$ in the overshoot regions.

The {double}-dot-dashed, green, curves in Fig.~\ref{f:punto} show the outcome for a 
computation in which the heavy-element abundance was reduced from the
reference value $Z=0.0295$, as determined from the LEGACY model fitting to the 
observed \Kepler frequencies (see Table~\ref{t:legacy-sample}), 
to the value $Z=0.018055$ {(and $Y=0.245$)}, adopted by the 3D simulation grid 
(see Section~\ref{s:3D-simulations}).
The comparison with the more metal-rich reference model (solid, black, curves) should
demonstrate the effect of our inconsistent use of heavy-element abundance between our
calibrated 1D stability computation and the 3D simulation grid. Although the effect 
of this inconsistency in $Z$ fortunately appears to be minute on $\eta$ and on the modal
surface corrections, it is our aim to recalibrate, in a future effort, consistently our 
1D models with 3D simulations of the appropriate composition, once such a 3D grid
for a range of metallicities will be available.

\subsection{Damping rates and modal surface effects}

The model results were obtained from calibrating the free convection parameters,
$a$, $b$, $c$, $\varPhi_{\rm s}$ and $\varPhi_{\rm c}$, together with the mixing-length
parameter $\alpha$, in a similar manner as discussed by \citet{AarslevEtal18}.
The parameters $c$, $\varPhi_{\rm s}$ and $\alpha$ are calibrated first, by iteration, 
to match the profiles of the 3D simulations for turbulent pressure, $\pt/p$, and 
anisotropy, $\varPhi$, in the mode-propagating layers, and to the surface 
convection-zone depths, $d_{\rm c}$, of the ASTFIT models. The remaining parameters,
$a$, $b$ and $\varPhi_{\rm c}$ are then determined by obtaining the best agreement between
theoretical damping rates and observed linewidths. The parameter $\varPhi_{\rm c}$
predominantly
affects the frequency of the depression in the damping rates, $\eta$, typically 
near $\nu_{\rm max}$, 
whereas the nonlocal parameter $a$ predominantly affects the depth of this depression 
in $\eta$. Finally, the parameter $b$ is calibrated 
by iteration, to reproduce the observed linewidths as well as possible at low and high
frequencies \citep[see also,][]{AarslevEtal18}. Owing to the rather low sensitivity of
$\eta$ to the parameter $b$, the value of $b$ resulting from this fit is somewhat uncertain.
 
The quality of the observed linewidths does not support a full
statistical analysis, including a formal $\chi^2$ minimization.
Moreover, the differences between theoretical damping rates and observed linewidths are 
still dominated by systematics brought about by our still poor understanding of modelling 
the convection-pulsation physics (see, for example, top left panel in Fig.~\ref{f:sun} for 
the solar case).
The aim of the procedure for finding the optimal model, as outlined above, is therefore
to find the best qualitative agreement with the 3D simulation results and 
linewidth measurements. 
In this fitting procedure we have, however, evaluated $\chi^2$ between observed linewidths
and damping rates for several of the parameter choices, 
and the parameter values listed in Table~\ref{t:legacy-sample} do
correspond to the fit with the lowest $\chi^2$ amongst the parameter sets
considered. This lowest $\chi^2$ between linewidths
and damping rates over the full observed frequency range is listed in
Table~\ref{t:legacy-results}.
Moreover, in addition to fitting the linewidths, our 1D models also aim to
simultaneously reproduce the turbulent pressure and anisotropy profiles of 3D simulations
in the mode-propagating layers
and the surface convection-zone depths, $d_{\rm c}$, of the ASTFIT models.

Detailed results for our sample of twelve LEGACY stars, listed in 
Table~\ref{t:legacy-sample}, are illustrated in Figs~\ref{f:LEGACY1}--\ref{f:LEGACY3}.
The left panels compare observed linewidths (symbols with error bars), which correspond 
to the full width at half maximum (FWHM) of the spectral peaks in the observed 
power spectrum, with
our calculated values of $\pi^{-1}\eta$ (solid curves, see caption of Fig.\ref{f:sun}) 
as a function of pulsation frequency. 
Also shown are the modal frequency corrections (dashed curve) relative to a standard
adiabatic oscillation calculation for the same equilibrium model, following the procedure
by \citet{HoudekEtal17}.
The right panels in Figs~\ref{f:LEGACY1}--\ref{f:LEGACY3} compare turbulent pressure
profiles, $p_{\rm t}/p$, and velocity anisotropies, $\varPhi$, between our 1D equilibrium
models (solid and dashed curves, respectively) and 3D simulations (dotted and dot-dashed
curves, respectively).
There is good agreement between the 3D-simulation results and calibrated 1D-equilibrium
models in the mode-propagating layers and, at the same time, also between the observed
linewidths and estimated values of $\pi^{-1}\eta$, for all considered LEGACY models. 
As in the solar case (Fig.~\ref{f:sun}) the stellar depths of the maxima 
of $p_{\rm t}/p$ in the 1D and 3D results
agree fairly well, except for models for which the heavy-element abundance $Z$ differs
considerably between our 1D computations and the 3D simulations, such as for the 
stellar models KIC~6116048 and KIC~9139163.
The effect of adopting the chemical abundances of the 3D simulations
(see Section~\ref{s:3D-simulations}) on the 1D-profiles of $p_{\rm t}/p$ are
indicated by the double-dot-dashed, green curves in Fig.~\ref{f:punto} for KIC~9139163, 
and in Fig.~\ref{f:LEGACY1} (bottom panels) for KIC~6116048. 
For KIC~6116048 and KIC~1435467 we adopt 
slightly larger values for $\varPhi_{\rm s}$, compared with the 3D results, leading to better
agreement between linewidth measurements and theoretical damping rates, though the effect
of varying $\varPhi_{\rm s}$ on $\pi^{-1}\eta$ is rather {small}.
Moreover, as discussed
also in Section~\ref{ss:varpar}, various 3D hydrodynamical simulation codes predict
noticeable differences for $\varPhi_{\rm s}$. 

\begin{figure}
	\includegraphics[width=\columnwidth]{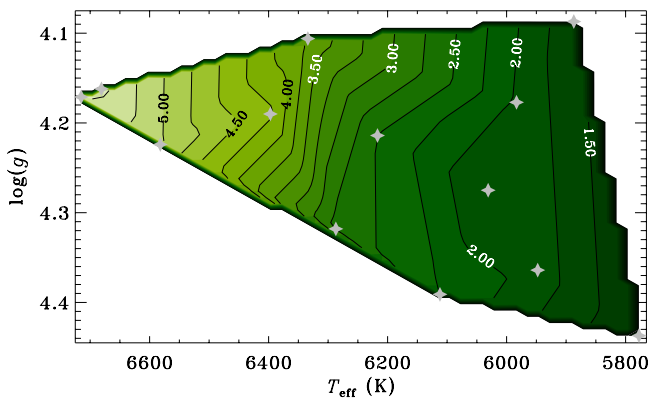}
    \caption{
     {Modelled $\pi^{-1}\eta$, where $\eta$ is the damping rate,} 
     corresponding to the 
     full width at half maximum (FWHM) of the spectral peaks in the power spectrum. 
     Results, in units of $\mu$Hz, are plotted as contours in a Kiel diagram at the
     observed frequency of maximum pulsation power, $\nu_{\rm max}$ 
     (see Table~\ref{t:legacy-results}).
     The star symbols indicate the locations of the twelve selected LEGACY stars and the Sun.     
     }
    \label{fig:damping-rates}
\end{figure}

\begin{figure}
	\includegraphics[width=\columnwidth]{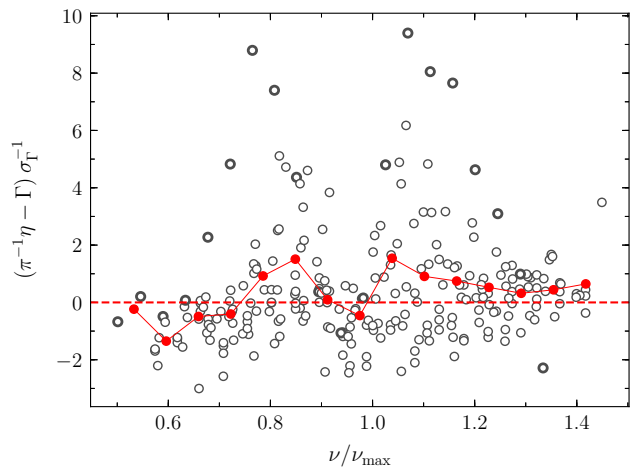}
    \caption{ Residuals between modelled $\pi^{-1}\eta$, 
             where $\eta$ is the damping rate,
             and observed linewidths, $\Gamma$, as a 
             function of the $\nu_{\rm max}$-normalized mode frequencies, 
             $\nu/\nu_{\rm max}$,
             {for all twelve LEGACY stars}. The residuals 
             have further been normalized by the uncertainty of the observed linewidths. 
             Each open circle corresponds to the value of a given pulsation mode.
             Red solid circles (connected) are uniformly weighted
             mean values of the individual mode values in 0.06 wide bins. The thick-edge
             open circles show the values for the Sun using BiSON mode linewidths.             
             }
    \label{fig:lwcomp}
\end{figure}

\begin{table*}
	\centering
	\caption{Solar and LEGACY model results at the frequency of maximum oscillation 
	power, $\nu_{\rm max}$. The first three columns are the Kepler Input Catalog (KIC)
	number, observed $\nu_{\rm max}$, and modal frequency corrections, $\delta\nu$. 
	The next two columns list the observed linewidths, $\Gamma$, and modelled $\pi^{-1}\eta$ 
	at $\nu_{\rm max}$. $\Gamma(\nu_{\rm max})$ was determined from fitting to the 
	data the analytical expression (1) by \citet{AppourchauxEtalCorr16}, 
	{and $\pi^{-1}\eta(\nu_{\rm max})$ was obtained 
	from linear interpolation in the modelled estimates $\eta(\nu)$}. 
    The last two columns show the reduced $\chi^2$ values between the observed, $\Gamma$, 
    and modelled, $\pi^{-1}\eta$, data, calculated as 
    $\chi^2_r= ({N-2})^{-1}\sum_{i=1}^N \sum_{j=1}^N (\Gamma_i - \pi^{-1}\eta_i)\, C^{-1}_{ij} \,(\Gamma_j - \pi^{-1}\eta_j)$, 
    where ${\bf C}$ is the covariance matrix for $\Gamma$, and $N$ 
    represents the number of fitted linewidths.
    Solar observations are from BiSON \citep{ChaplinEtal05} and solar model results
    from \citet{HoudekEtal17}.
    {For the BiSON linewidths we adopted a diagonal covariance matrix.}
	}
	\label{t:legacy-results}
	\begin{tabular}{rcrccrc}        
		\hline
		KIC & $\nu_{\rm max}\      (\mu$Hz) & $\delta\nu\    (\mu$Hz) 
		    & $\Gamma\             (\mu$Hz) & $\pi^{-1}\eta\ (\mu$Hz) 
		    & $\chi^2$ & $N$\\
		\hline
		Sun     & 3090 &10.0 & 1.06$\,\pm\,$0.03 & 1.25 & 25.20 & 20\\
		6933899 & 1390 & 3.7 & 1.30$\,\pm\,$0.07 & 1.56 & 10.92 & 13\\
		10079226& 2653 & 7.4 & 2.05$\,\pm\,$0.35 & 1.89 &  3.81 & 12\\
		10516096& 1690 & 4.6 & 1.56$\,\pm\,$0.10 & 1.97 & 12.33 & 13\\
		6116048 & 2127 & 6.2 & 1.62$\,\pm\,$0.09 & 1.73 & 17.74 & 15\\
		8379927 & 2795 & 9.2 & 2.43$\,\pm\,$0.11 & 2.25 & 11.38 & 17\\
		12009504& 1866 & 5.9 & 2.38$\,\pm\,$0.14 & 2.49 &  4.98 & 15\\
		6225718 & 2364 & 8.1 & 2.58$\,\pm\,$0.12 & 2.66 &  7.37 & 19\\
		1435467 & 1407 & 7.0 & 5.18$\,\pm\,$0.17 & 3.70 &  4.13 & 15\\
		9139163 & 1730 & 7.9 & 5.28$\,\pm\,$0.13 & 4.30 &  3.41 & 19\\
		9206432 & 1866 & 8.7 & 5.87$\,\pm\,$0.27 & 5.02 &  1.90 & 16\\
		2837475 & 1558 &10.1 & 6.39$\,\pm\,$0.20 & 5.57 &  1.91 & 18\\
		11253226& 1591 & 9.5 & 5.80$\,\pm\,$0.17 & 5.53 &  5.26 & 19\\
		\hline
	\end{tabular}
\end{table*}

Values of the LEGACY linewidths, $\Gamma$, and computed $\pi^{-1}\eta$ are listed
in Table~\ref{t:legacy-results} at the frequency of maximum oscillation power, 
$\nu_{\rm max}$, together with our modal frequency corrections, 
$\delta\nu(\nu_{\rm max})$.
The values for $\Gamma$ 
are obtained from fitting
Appourchaux's\,et\,al.\,(\citeyear{AppourchauxEtalCorr16}) expression~(1) to the 
observed {linewidths}, following the procedure by \citet{LundEtal17},
and the $\pi^{-1}\eta$ values are obtained from interpolating in the frequency-dependent
damping rates.

As reported before \citep{ChaplinEtal09, BelkacemEtal12}, the increase of
the observed linewidths with increasing surface temperature, $T_{\rm eff}$, is
also reproduced here with our stability analysis, as illustrated in 
Fig.~\ref{fig:damping-rates}. There is also a dependence 
on surface gravity \citep[see also][]{Houdek17}, but of much smaller magnitude.
Moreover, the decrease of the amount of the depression in the linewidths near 
$\nu_{\rm max}$ with increasing $T_{\rm eff}$, as
reported previously by \citet{AppourchauxEtal14} and \citet{LundEtal17}, is also 
reproduced by our theoretical damping rates (see Figs~\ref{f:LEGACY1}--\ref{f:LEGACY3}).

Fig.~\ref{fig:lwcomp} shows the residuals between observed linewidths and 
estimated $\pi^{-1}\eta$ for all fitted modes,  divided 
by the measurement uncertainties, and on a frequency scale normalized by $\nu_{\rm max}$.
There is still a discrepancy between observations and models just below 
and above $\nu_{\rm max}$, as can also be seen for the solar case 
(upper left panel of Fig.~\ref{f:sun}). 
This discrepancy is predominantly a consequence of the
still incompletely modelled superadiabatic boundary layers. The thermal relaxation time
of the superadiabatic boundary layer, when expressed as a frequency, $\nu_{\rm p}$, 
is of similar order as the frequency at which the depression in the linewidths 
is observed ($\nu_{\rm p}\simeq$ 2.8\,mHz in the solar case, the exact value depending upon 
the nonlocal convection parameters; \citealt{Balmforth92a}, \citealt{ChaplinEtal05}). 
It is at this
frequency $\nu_{\rm p}$ where the energy exchange between the pulsation and the stellar
background is most efficient \citep[e.g.,][]{HoudekDupret15}.
Any insufficient representation of either the stellar mean stratification or the
thermal and dynamical properties of the super\-adiabatic boundary layers are revealed by 
a mismatch between the modelled damping rates and linewidth measurements.
In a future investigation we plan to identify the reason for this mismatch by
comparing the properties of the super\-adiabatic boundary layer with 3D simulation results.
It is interesting to note that a more careful calibration of the nonlocal convection
parameter $b$ (see Section~\ref{s:stability}) with the help of 3D simulation results,
together with an analysis of work integrals, could identify possible missing physics 
in our model calculations, such as the effect of kinetic energy transport 
\citep[e.g.,][]{Gough12}.

\begin{figure}
	\includegraphics[width=\columnwidth]{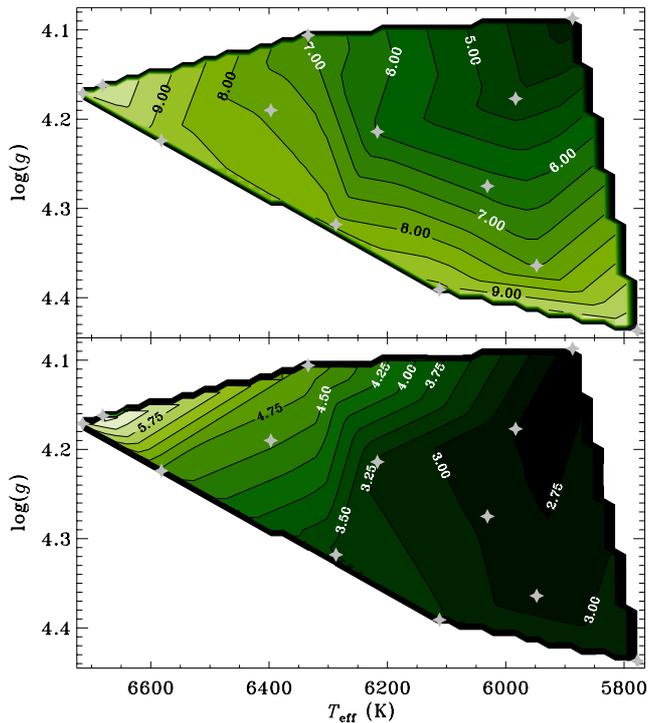}
    \caption{
     Modal frequency corrections, $\updelta\nu$, relative to a standard adiabatic
     pulsation calculation. Results are plotted as contours in a Kiel diagram at the
     observed frequency, $\nu_{\rm max}$ 
     (see Table~\ref{t:legacy-results}), 
     of maximum pulsation power.
     Top: the contours are labelled in units of $\mu$Hz.
     Bottom: contours show $(\updelta\nu/\nu_{\rm max})\times10^{3}$  
     in order to illustrate the relative modal frequency shift with respect to
     mode frequency.
     The star symbols indicate the locations of the twelve selected LEGACY stars and the Sun.
     }
    \label{fig:nu-corr}
\end{figure}

In Fig.~\ref{fig:nu-corr} the modal frequency corrections, $\delta\nu$, at $\nu_{\rm max}$,
also listed in Table~\ref{t:legacy-results}, are plotted as contours
in a Kiel diagram.
The top panel shows $\delta\nu$ in $\mu$Hz, whereas the bottom panel
displays the relative modal correction, $\delta\nu/\nu_{\rm max}$, in order 
to show the relative importance of the modal frequency shift with respect to
mode frequency.
There is a clear trend with stellar surface values, 
particularly the increase of $\delta\nu/\nu_{\rm max}$
with surface temperature.

\section{Discussion and conclusions}
\label{s:discussion}

It has been the aim of this paper to demonstrate the simultaneous 
modelling of observed and 3D-simulated stellar properties 
in main-sequence 
stars that are typically not included in standard stellar evolutionary calculations. 
In particular, we included in our 1D stellar models turbulent pressure
and velocity-anisotropy profiles as suggested by realistic 3D simulations of convective
stellar atmospheres, 
and, at the same time, obtained good agreement between pulsation-linewidth measurements
and modelled damping rates over the full observed frequency range.
The remaining differences, e.g., in the $\pt/p$-profiles between 1D and 3D models in the
deeper, mode-propagating layers, help us to assess missing physics, 
such as the kinetic energy flux \citep[e.g., ][]{Gough12}, convective backwarming
\citep[e.g.,][]{TrampedachEtal17} and the asymmetry between up- and downflows, that is 
required for further improving our current 1D models for convective energy transport.

It is interesting to note that models with deeper surface convection zones, compared with the 
ASTFIT models (see Section~\ref{s:legacy-sample}),
reproduce the observed pulsation-linewidth profiles somewhat better, particularly 
for the hotter stars in our LEGACY stellar sample (see for example Fig.~\ref{f:punto}).
This is a result of the
increasing damping rates with increasing mixing length, mostly due to the contributions
from the turbulent pressure perturbations.
This effect may therefore serve as an additional constraint for calibrating convection 
in stellar models and the consequent extent of surface convection zones. 
Other procedures to constrain surface convection-zone depths include the use of acoustic 
glitches \citep[e.g.,][]{Gough90}.
This method, however, needs very accurate pulsation frequencies, and current glitch 
models are still affected by the poorly understood glitch contribution from hydrogen
ionization and the difficult-to-determine location of the star's acoustic surface
\citep[e.g.,][]{HoudekGough07, MazumdarEtal14, ReeseEtal16}.

Another interesting outcome of our stability analyses is the prediction of the
modal frequency contribution to the so-called `surface effects'. By combining our modal
frequency corrections, $\delta\nu$, relative to a standard adiabatic 
pulsation calculation,
with structural, adiabatic, frequency 
corrections, $\delta\nu^{\rm s}$ \citep[e.g.,][]{TrampedachEtal17}, 
we would have, for the first time, a purely physical description for the total 
surface frequency corrections, $\Delta\nu:=\delta\nu+\delta\nu^{\rm s}$, at hand. 
For a calibrated solar model, \citet{HoudekEtal17} obtained at the frequency of
maximum oscillation power, $\nu_{\rm max}\simeq3090\,\mu$Hz, the values
$\delta\nu\simeq10.0\,\mu$Hz and $\delta\nu^{\rm s}\simeq-16.0\,\mu$Hz, 
leading to an underestimation (i.e. over-correction) of the fully surface-corrected 
model frequency\footnote{The standard solar model \citep{JCDEtal96} 
overestimates the adiabatically 
computed frequency by $\sim4.6\,\mu$Hz at $\nu_{\rm max}$, relative 
to the observations.} 
of $\sim1.4\,\mu$Hz, relative to the observed solar frequency
(see fig.~5 of \citealt{HoudekEtal17}).
We have, however, to remain aware how stellar-model fits to
seismic observations, such as the ASTFIT models (Section~\ref{s:legacy-sample}), exhibit 
coupled parameters. In particular
the surface effect, mixing length (e.g., \citealt{LiEtal18}) and helium abundance
can be strongly correlated, stressing the importance of constraining these quantities
independently. Moreover, $\delta\nu$ also depends on metallicity as demonstrated 
in Fig.~\ref{f:punto} for KIC~9139163. 
We therefore plan to address these open issues in future work by exploiting additional
information from 3D simulation results in a fully consistent way, such as adopting 
the 3D atmosphere structures \citep{TrampedachEtal14a} and 3D-simulation-calibrated 
mixing lengths \citep{TrampedachEtal14b} in the ASTFIT models 
(Section~\ref{s:legacy-sample}). 
Also, a necessary step will be the implementation of a pipeline that will take into
account a full statistical analysis of the measured linewidth errors and a
consistent treatment of frequency corrections, based on our theoretical formulation
involving a physical model, in
both the ASTFIT and pulsation-stability calculations. 
 
We have demonstrated that the additional stellar properties, considered in this work,
further constrain the theory of stellar structure and
evolution. The combined effort of improving the modelling in this manner and extending
the analysis to a broader range of observed stars will, one
may hope, improve our understanding of the properties of stellar
convection and its interaction with pulsations. Such improvements
in the modelling of the near-surface layers are also crucial
for reducing their effect on the computed frequencies and consequently
improving the characterization of stellar properties through
frequency fitting.

\section*{Acknowledgements}
We thank Douglas Gough for many inspiring discussions.
We also thank the referees for helpful comments.
MNL acknowledge the support of the ESA PRODEX programme and The Danish Council for
Independent Research | Natural Science (Grant DFF-4181-00415).
RT acknowledges funding from NASA grant {80NSSC18K0559}.
Funding for the Stellar Astrophysics Centre is provided
by The Danish National Research Foundation (Grant DNRF106).




\bibliographystyle{mnras}
\bibliography{ghrefs} 




\appendix

\section{Detailed results for all twelve calibrated LEGACY models}
In this appendix we plot the frequency-dependent damping rates and modal surface corrections 
for our sample of twelve LEGACY stars (see Table~\ref{t:legacy-sample}), together with
the calibrated 1D and 3D solutions of the turbulent pressure profiles, $\pt$, and velocity 
anisotropies, $\varPhi$.
 


\begin{figure*}
	\includegraphics[width=0.98\textwidth]{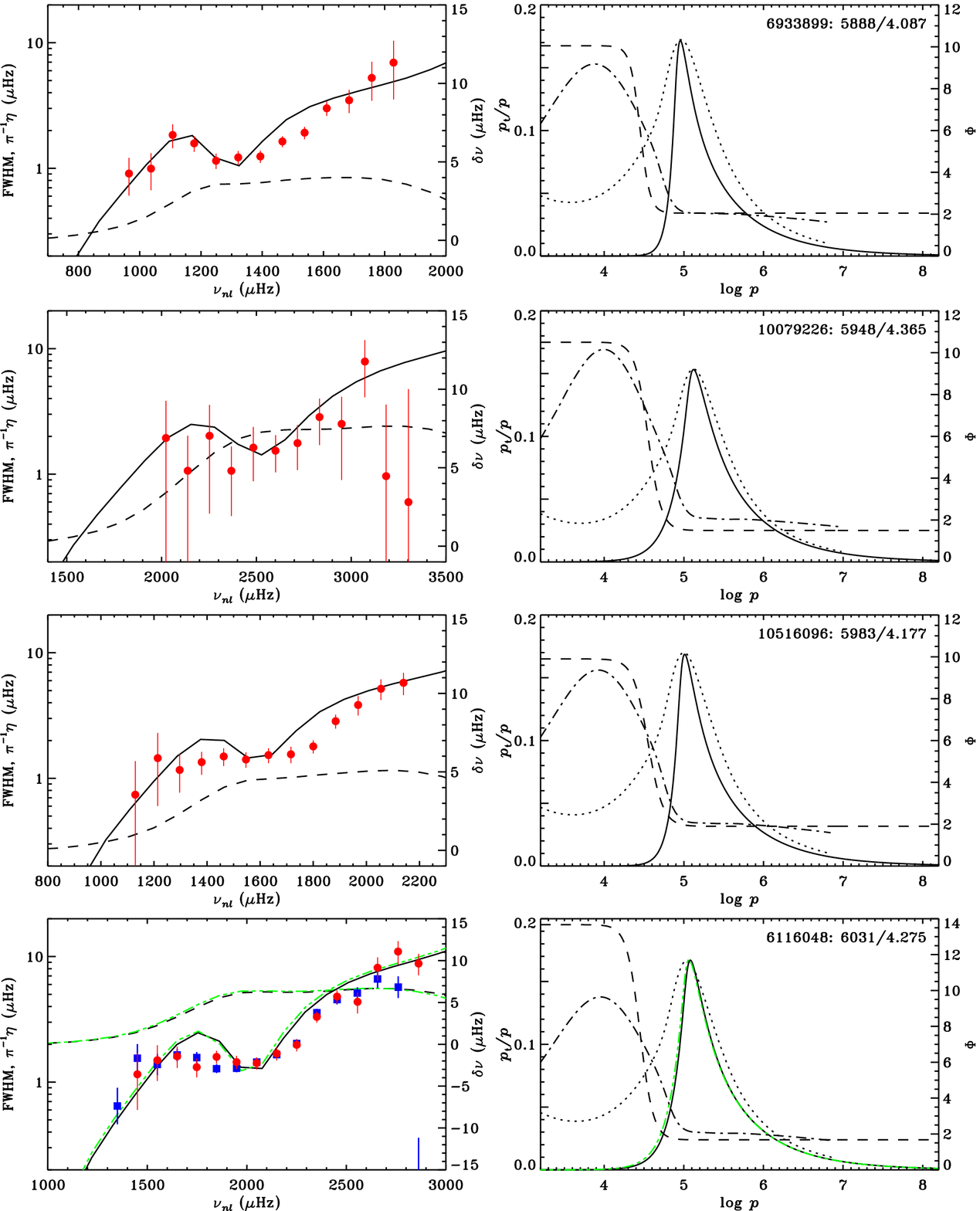}
    \caption{LEGACY model results. The KIC numbers and surface values, 
    $T_{\rm eff} (K)\,$/$\,\log\cg$, of the LEGACY stars are indicated in the right panels.
    Plots are arranged in order of increasing $T_{\rm eff}$ from top to bottom.
    Left: comparison between observed linewidths   
    (FWHM, red solid circles with error bars are the LEGACY data from \citealt{LundEtal17}, 
    blue filled squares with error bars are from \citealt{AppourchauxEtal14, 
    AppourchauxEtalCorr16}) 
    and $\pi^{-1} \eta$ (see caption to Fig.~\ref{f:sun}).
    The dashed curves are the 
    radial modal frequency corrections, $\updelta\nu$, relative to a standard adiabatic
     pulsation calculation. Right: comparison of the turbulent pressure profile 
    $p_{\rm t}/p$ between 3D simulations (dotted curves; \citealt{TrampedachEtal14a})
    and 1D nonlocal envelope model (solid curves). The dot-dashed curves are 
    the turbulent velocity anisotropies $\varPhi$
    of 3D simulations and the dashed curves the adopted depth-dependent velocity
     anisotropies in the 1D models.
     The double-dot-dashed, green curves for KIC~6116048 adopt the 3D-grid abundances,
      $Z=0.018055$ and $Y=0.245$.
    } 
    \label{f:LEGACY1}
\end{figure*}

\begin{figure*}
	\includegraphics[width=0.98\textwidth]{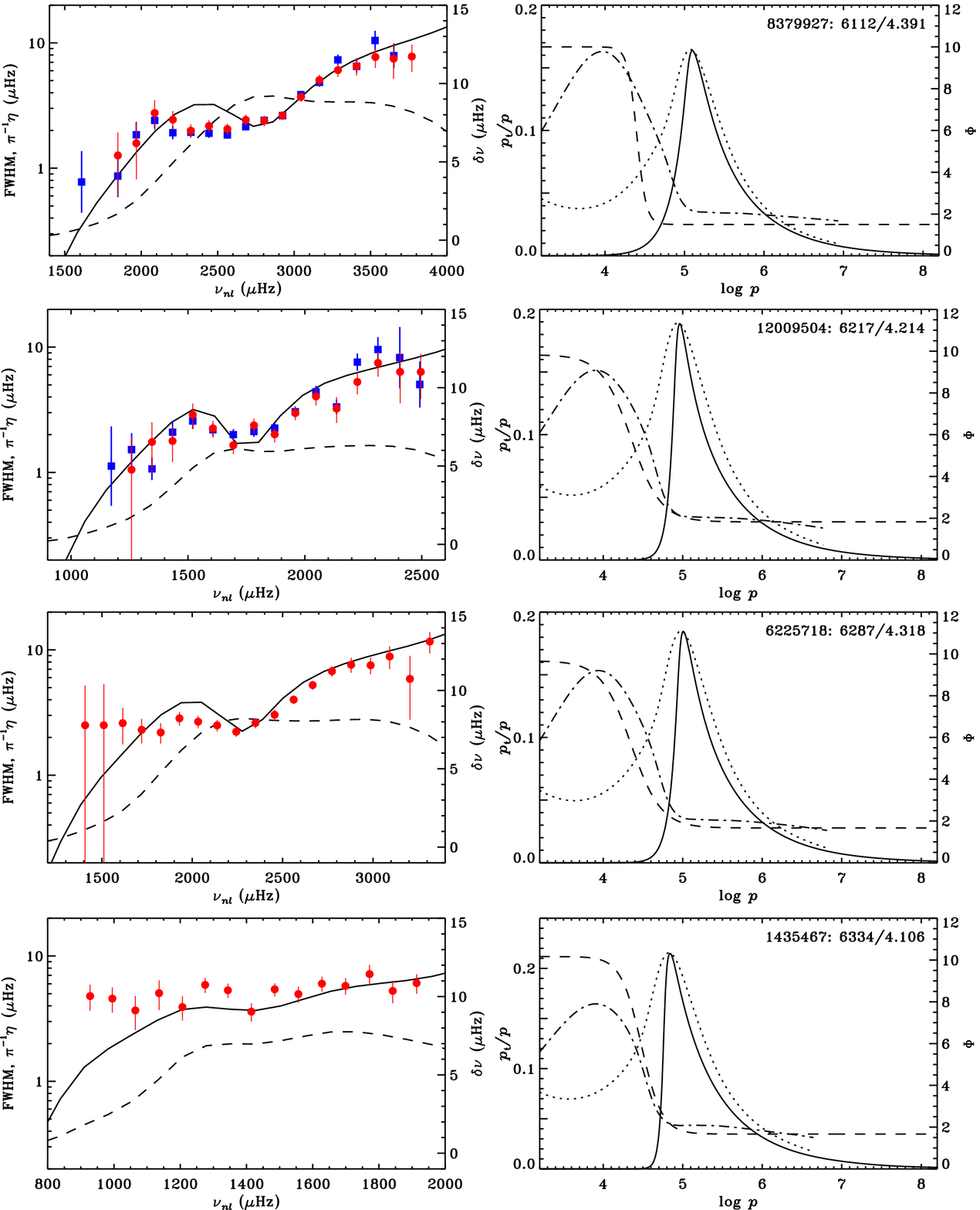}
    \caption{LEGACY model results. Details are as in the caption of Fig.~\ref{f:LEGACY1}.
    } 
    \label{f:LEGACY2}
\end{figure*}

\begin{figure*}
	\includegraphics[width=0.98\textwidth]{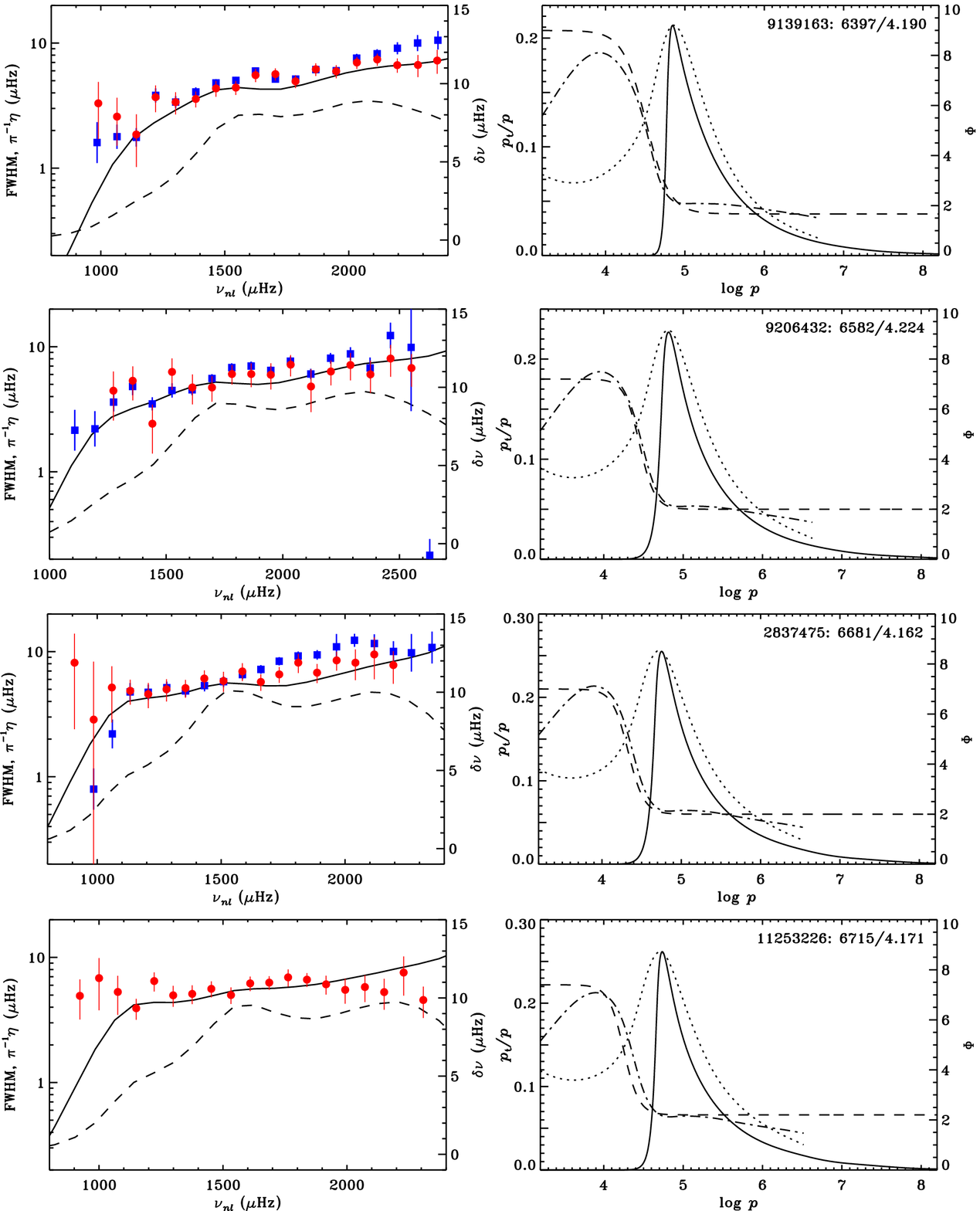}
    \caption{LEGACY model results. Details are as in the caption of Fig.~\ref{f:LEGACY1}.
    } 
    \label{f:LEGACY3}
\end{figure*}

\bsp	
\label{lastpage}
\end{document}